\documentclass{aa}  

\usepackage{graphicx}
\usepackage{xcolor}
\usepackage{txfonts}
\usepackage[colorlinks,linkcolor={blue},citecolor={blue}]{hyperref}
\begin{document} 

   \title{Theoretical wind clumping predictions of OB supergiants from line-driven instability simulations across the bi-stability jump}
  
  \titlerunning{Clumping in OB supergiant winds}
  
   \author{F.~A.~Driessen,
          J.~O.~Sundqvist,
          \and
          N.~D.~Kee
          }

   \institute{Institute of Astronomy, KU Leuven,
              Celestijnenlaan 200D box 2401, 3001, Leuven, Belgium\\
              \email{florian.driessen@kuleuven.be}
             }

   \date{Received ....; accepted ...}

% \abstract{}{}{}{}{} 
% 5 {} token are mandatory
 
  \abstract
   {The behaviour of mass loss across the so-called `bi-stability jump' -- where iron recombines from Fe IV to Fe III -- is a key uncertainty in models of massive stars. Namely, while an increase in mass loss is theoretically predicted, this has so far not been observationally confirmed. However, radiation-driven winds of hot, massive stars are known to exhibit clumpy structures triggered by the line-deshadowing instability (LDI). Such wind clumping severely affects empirical mass-loss rates inferred from $\rho^2$-dependent spectral diagnostics. As such, if clumping properties differ significantly for O and B supergiants across the bi-stability jump, this may help alleviate current discrepancies between theory and observations.}   
   {We investigate with analytical and numerical tools how the onset of clumpy structures behave in the winds of O supergiants (OSG) and B supergiants (BSG) across the bi-stability jump.}
   {We derive a scaling relation for the linear growth rate of the LDI for a single optically thick line and apply it in the OSG and BSG regime. We run 1D time-dependent line-driven instability simulations to study the non-linear evolution of the LDI in clumpy OSG and BSG winds.}
   {Linear perturbation analysis for a single line shows that the LDI linear growth rate $\Omega$ scales strongly with stellar effective temperature and terminal wind speed: $\Omega \propto \varv_\infty^2 T_{\mathrm{eff}}^4$. This implies significantly lower growth rates for (the cooler and slower) BSG winds than for OSG winds. This is confirmed by the non-linear simulations, which show significant differences in OSG and BSG wind structure formation, with the latter characterized by significantly weaker clumping factors and lower velocity dispersions. This suggests that, indeed, lower correction factors due to clumping should be employed when deriving empirical mass-loss rates for BSGs on the cool side of the bi-stability jump. Moreover, the non-linear simulations provide a theoretical background toward explaining the general lack of observed intrinsic X-ray emission in (single) B star winds.}
   {}


   \maketitle
%
%-------------------------------------------------------------------

\section{Introduction}
 
Determining and understanding mass-loss rates of massive stars is of great importance because these stars' lives are heavily dependent on the amount of mass they lose throughout their lifetime \citep{1986ARA&A..24..329C}. The mass expelled from the surface of hot, massive OB stars is known to be driven by the scattered radiation off spectral lines \citep[][hereafter CAK]{1975ApJ...195..157C} via a line-driven wind. Subsequent refinements to CAK theory \citep{1986ApJ...311..701F,1986A&A...164...86P} provided a basic quantitative understanding of the global properties of these line-driven winds, such as the metallicity dependence of mass-loss rate \citep{2001A&A...369..574V,2007A&A...473..603M} and the relation between the stellar luminosity and wind momentum \citep{1995svlt.conf..246K}. 

Nowadays mass-loss rates of OB stars are either calculated from theoretical line-driven wind models (more below) or inferred from observational spectral diagnostics. A key problem regarding theory and observation is the so-called \emph{bi-stability jump} \citep{1990A&A...237..409P} in the effective temperature range of 22.5 - 27.5 kK. Physically at these temperatures iron recombines to a lower ionization state (Fe IV $\rightarrow$ Fe III) and has then more driving lines leading to a stronger radiation force. Across the bi-stability jump \citet{1995ApJ...455..269L} observed a drop in terminal wind speed of the wind. However, line-driven wind models \citep{1999A&A...350..181V} predict an additional increase in mass-loss rate by a factor of several when transiting from the hot to cool side of the bi-stability jump. Whereas a decrease in terminal wind speed is observationally established \citep{2007A&A...467.1265B,2008A&A...478..823M,2018A&A...614A..91H}, the predicted increase in mass-loss rate has not been confirmed. Such a lack of mass loss increase over the bi-stability jump might also have rather severe consequences for massive star evolution studies \citep{2010A&A...512L...7V}. Recently, \citet{2017A&A...598A...4K} concluded that without a bi-stability mass loss increase, massive stars do not spin down fast enough to explain the rather large observed population of slowly rotating B supergiants. 

However, there might be a bias regarding empirical mass-loss rates derived from observations across the bi-stability jump. Namely, nowadays there is ample evidence that line-driven winds are both spatially and temporally structured and variable, i.e.~they are inhomogeneous, \emph{clumpy} winds \citep[see, e.g.][for reviews]{2008A&ARv..16..209P,2011A&A...528A..64S}. Such wind clumping can then seriously affect empirical mass-loss rates derived from observations. In particular, spectral diagnostics of $\rho^2$-dependent atomic processes are affected by wind clumps (e.g. the optical recombination line H$\alpha$). To see this, consider how small scale density inhomogeneities redistribute matter into overdense clumps and an effective void interclump medium. Using conservation of mass, the overall average density in the wind scales with the mass-loss rate as $\langle \rho \rangle \propto \dot{M}$. The density of the clumps can then be written as $\rho_{\mathrm{clump}} = \langle \rho \rangle/f_{\mathrm{vol}}$ with $f_{\mathrm{vol}}$ accounting for the clump volume. Similarly, it follows $\rho_{\mathrm{clump}}^2 = \langle \rho^2 \rangle/f_{\mathrm{vol}}$ . To quantify the wind clumping for these $\rho^2$-dependent diagnostics, it is common to define a clumping factor $f_{\mathrm{cl}} \equiv \langle \rho^2 \rangle/\langle \rho \rangle^2 = 1/f_{\mathrm{vol}} \geq 1$. In the case of recombinations between an electron-proton pair, the opacity $\chi$ can be related to the mass-loss rate via $\chi \propto \langle \rho^2 \rangle = \langle \rho \rangle^2 f_{\mathrm{cl}} \propto \dot{M}^2 f_{\mathrm{cl}}$. For the important H$\alpha$ mass loss diagnostic this means then that wind clumping leads to an overestimate of mass-loss rates with a $\sqrt{f_{\mathrm{cl}}} $ dependency, if it is not properly accounted for in the diagnostic modelling, i.e.~$\dot{M}_{\mathrm{unclumped}} = \dot{M}_{\mathrm{clumped}}\sqrt{f_{\mathrm{cl}}}$ \citep{2006A&A...454..625P,2018A&A...619A..59S}. This effect could explain the theoretically predicted increase in mass-loss rate at the bi-stability jump, if clumping factors change significantly across the temperature region where this occurs. A first clue that this might be true has been provided by \citet{2014A&A...565A..62P}, who found that H$\alpha$ line-profiles switch from strong to weak emission across the bi-stability jump. Inferring wind clumping properties is, however, a non-trivial task, and can be either done by a multitude of empirical spectral wind diagnostics \citep[e.g.][]{2006A&A...454..625P} or by performing theoretical complex time-dependent non-linear hydrodynamic simulations (as presented here). 

These hydrodynamic simulations are motivated by linear stability analysis that has shown that line-driving of hot star winds is in fact highly unstable due to a very strong, intrinsic \emph{line-deshadowing instability} (LDI) occurring on small spatial scales \citep{1979ApJ...231..514M,1980ApJ...241.1131C,1984ApJ...284..337O}. Indeed, subsequent time-dependent line-driven instability models \citep{1988ApJ...335..914O,1995A&A...299..523F,2003A&A...406L...1D,2018A&A...611A..17S} have confirmed the picture of a highly inhomogeneous, \emph{clumpy} wind. Such small-scale structured, clumpy winds also provide basic explanations of key observed signatures in the winds of massive stars, e.g.~soft X-ray emission \citep{1997A&A...322..167B,2010MNRAS.405.2391C,2017SSRv..212...59M}, extended regions of zero residual flux in some UV resonance lines \citep{1983ApJ...274..372L,2010A&A...510A..11S}, or migrating spectral subpeaks in optical recombination lines \citep{1998ApJ...494..799E,2005A&A...432..281D}. 

A key point, however, is that all these previous LDI simulations have been carried out for stellar parameters typical for O supergiant winds. So far, no LDI model has been calculated for a B supergiant. The bi-stability jump happens where spectroscopic classification approximately separates the O and B star regime, requiring new LDI models for the latter. These models provide a first step toward a better characterization of clumping properties and mass loss for B supergiants. In turn, this characterization of clumping and mass loss might then also provide better constraints regarding these objects' rather poorly understood evolutionary status \citep{2012ARA&A..50..107L}.   

Inspired by this, we perform time-dependent line-driven instability simulations of both O and B supergiant winds. We carefully examine and contrast their clumping behaviour and discuss the implications. By means of analytical perturbation analysis we illustrate the difference in linear growth rate of the LDI between O and B supergiant winds. These analytical calculations are verified with non-linear, numerical simulations of the LDI. We test whether wind clumping in OB supergiant winds undergoes a structural change across the bi-stability jump. Specifically, we look at wind clumping in the H$\alpha$ line-formation region in the inner wind because of the importance of H$\alpha$ as mass-loss rate diagnostic. 

%--------------------------------------------------------------------
\section{Scaling relation for the growth rate of unstable structures using perturbation analysis}

Following \citet{1984ApJ...284..337O} we consider small-amplitude perturbations in velocity $\delta \varv \propto e^{i(kx-\omega t)}$ applied to the unpertubed 1D momentum equation. Here $k$ denotes the wavenumber and $\omega$ the angular frequency that can be either real, imaginary, or complex. For a spherically symmetric flow locally co-moving with the mean flow, and in the limit of vanishing gas pressure, the first-order perturbed momentum equation becomes
\begin{equation}\label{eq:perturbedeom}
\omega = i\frac{\delta g}{\delta \varv},
\end{equation}

\noindent with $\delta g$ the response of the mean line-force to the perturbed velocity. By considering an optically thick line in the mean flow and perturbations in wavelength and optical depth, \citet{1984ApJ...284..337O} derived a general \emph{bridging law} for the perturbed line-force
\begin{equation}
\frac{\delta g}{\delta \varv} \approx \Omega \frac{ik\Lambda}{1 + ik\Lambda},
\end{equation}

\noindent with $\Lambda$ the bridging length which is on the order of a radial Sobolev length $L\equiv \varv_{\mathrm{th}}/(d\varv /dr) \sim \Lambda$ while $\Omega\approx g_{\mathrm{Sob}}/\varv_{\mathrm{th}}$ the growth rate of the instability, $g_{\mathrm{Sob}}$ the Sobolev line-force, and $\varv_{\mathrm{th}}$ the thermal speed in the wind. Note that the perturbed quantities $\delta g$, $\delta \varv$ are taken with respect to the background unperturbed, steady-state mean flow described by Sobolev theory for a single optically thick line. Mathematically this gives rise to the Sobolev dependency of the bridging length and growth rate of the instability, albeit the instability itself is of non-Sobolev nature.

It follows that long-wavelength perturbations ($k\Lambda \ll 1$) lead to $\mathcal{R}[\omega]$ resulting in stable, radiative-acoustic waves \citep{1980ApJ...242.1183A} while in the limit of short-wavelength perturbations ($k\Lambda \gg 1$) we retrieve $\mathcal{I}[\omega]$ meaning an instability occurs \citep{1979ApJ...231..514M,1980ApJ...241.1131C}. This instability, the line-deshadowing instability, is very strong and inherent to a line-driven wind.

The above bridging law is valid for pure photon absorption, i.e.~perturbations in the direct line-force component. Inclusion of line-scattering, captured by the perturbed diffuse line-force, results in a reduction of the growth rate near the stellar base \citep{1985ApJ...299..265O}. Since line-scattering is part of our hydrodynamical models (detailed in Section \ref{sec:hydro}), following \citet{1996ApJ...462..894O} (hereafter OP96), a scattering-modified bridging law is
\begin{equation}
\frac{\delta g}{\delta \varv} \approx \Omega \frac{kL}{1+(kL/2x_\star)^2} \left[ (1-s)\frac{kL}{2x_\star}\right],
\end{equation}

\noindent with $x_\star \approx 1$ the frequency at the blue-absorption edge (see OP96 for definition), and
 \begin{displaymath}
      s      = \frac{1}{1+\mu_\star}    \, , \;
      \mu_\star = \sqrt{1-\left( \frac{R_\star}{r} \right)^2}.
\end{displaymath}

\noindent Taking now the limit of short-wavelength, unstable perturbations ($kL\gg 1$) this becomes
\begin{equation}
\frac{\delta g}{\delta \varv} \approx 2\Omega(1-s),
\end{equation}

\noindent A simple scaling relation for the above perturbed line-force then follows from applying Sobolev theory in the optically thick limit for a single line
\begin{equation}
\frac{g_{\mathrm{Sob}}}{\varv_{\mathrm{th}}} \propto \frac{L_\star}{r^2} \frac{(d\varv /dr)/\rho}{\varv_{\mathrm{th}}}.
\end{equation}

\noindent {Recalling that $\dot{M} \propto r^2\rho \varv$ and considering the maximum growth rate in the limit far away from the star ($s\rightarrow 0.5$, i.e.~the supersonic regime in which the Sobolev approximation holds \citep{1960mes..book.....S}: $\varv \gg \varv_{\mathrm{th}}$)
\begin{equation}\label{eq:scalingrel}
\begin{split}
\frac{\delta g}{\delta \varv} \approx \Omega &\propto \frac{\varv_\infty^2}{\varv_{\mathrm{th}}}\frac{L_\star}{R_\star}\frac{1}{\dot{M}}, \\
&\propto \varv_\infty \left( \frac{\varv_\infty}{\varv_{\mathrm{th}}} \right) \frac{T_{\mathrm{eff}}^4 R_\star}{\dot{M}},
\end{split}
\end{equation}

\noindent where we have used $L_\star \propto R_\star^2 T_{\mathrm{eff}}^4$ in the last expression. A basic dependency of $\varv_\infty / \varv_{\mathrm{th}}$ appears in this relation that has also been found in the analysis of \citet{1984ApJ...284..337O}. The latter considered a scaling of the growth rate compared to the typical wind expansion rate by approximating $\varv(d\varv/dr) \sim g_{\mathrm{CAK}}$, with $ g_{\mathrm{CAK}}$ the total mean CAK line-force.

If we plot this maximum $\Omega$ dependency for our models it can be clearly seen that the relative growth rate of the LDI differs significantly between O and B supergiants (Figure \ref{FIGgrowthrates}). Between the hottest and coolest model there is a decrease in growth rate of about 90\%. From our small-amplitude Ansatz the amplitude of the instability grows with $\exp{\Omega t}$ with $\Omega^{-1}$ a characteristic time-scale (growth time). As such it can be understood that $\Omega^{-1}$ denotes then the characteristic time at which the wave gets amplified by a factor $e$, providing a relevant time-scale of the problem. 
   
From this analysis it follows then that the onset of structure formation in O supergiants should occur much faster than in B supergiants, and that the resulting structure should also be more vigorous in O supergiants.  Indeed, as shown further below this is exactly what we observe in our non-linear, numerical simulations.  
   
      \begin{figure}[ht!]
   \centering
   \includegraphics[width=\hsize]{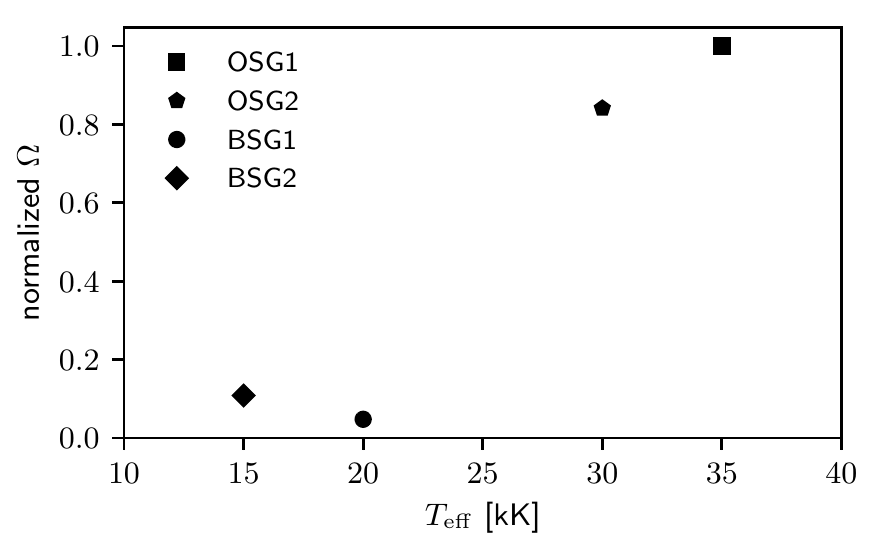}
      \caption{Comparing maximum growth rates for OB supergiants as function of the effective temperature using the scaling relation from equation \eqref{eq:scalingrel}. All models are normalized with respect to the hottest model to show the relative difference. Parameters used as listed in Table \ref{table:params}.}
         \label{FIGgrowthrates}
   \end{figure}  

%--------------------------------------------------------------------
\section{Numerical simulations of clumpy winds}\label{sec:hydro}

\begin{table*}
\caption{Input stellar and wind parameters of our models. For comparison we include mass-loss rates as predicted by \citet{2001A&A...369..574V} using the ratios $\varv_\infty / \varv_{\mathrm{esc_eff}}=1.3,2.6$ for below and above the bi-stability, respectively.}             
\label{table:params}      
\centering          
\begin{tabular}{c c c c c c c c c c c c c c}     
\hline\hline       
                      % To combine 4 columns into a single one 
Model & $L_\star$ & $M_\star$  & $R_\star$ & log $g$ & $T_{\mathrm{eff}}$ & $\alpha$ & $\bar{Q}$ & $Q_{\mathrm{max}}$ & $\varv_{\mathrm{esc},\mathrm{eff}}$  & $\dot{M}_{\mathrm{Vink}}$ \\ 

           & [$L_\odot$] & [$M_\odot$] &  [$R_\odot$] & [g cm$^{-3}$] & [kK] & &  & [$\bar{Q}$] & [km s$^{-1}$] & [$M_\odot$ yr$^{-1}$] \\
\hline                    
   OSG1 & $6\times 10^{5}$ & 40 & 20 & 3.43 & 35 & 0.6 & 2000 & 0.004  & 678 &$4.1\times 10^{-6}$ \\ 
   OSG2 & $4\times 10^{5}$ & 30 & 20 & 3.31 & 30 & 0.6 & 2000 & 0.004 & 608 &$1.5\times 10^{-6}$\\
   BSG1  & $1\times 10^{5}$&  20 & 40 & 2.53 & 20 & 0.5 & 2000 & 0.004  & 407 & $7.0\times 10^{-7}$\\
   BSG2 & $6\times 10^{4}$ & 20 & 40 & 2.53 & 15 & 0.5 & 2000 & 0.004 & 419  & $3.2\times 10^{-7}$  \\
\hline                  
\end{tabular}
\end{table*}

\subsection{Hydrodynamics}

To perform simulations of the non-linear evolution of the radiatively unstable stellar wind we evolve the equations of 1D spherically symmetric isothermal radiation-hydrodynamics, assuming the wind is at the same temperature as the stellar effective temperature. Table \ref{table:params} shows the adopted stellar and wind parameters over a range of representative effective temperatures for OB supergiants in the Galaxy (see Section \ref{sec:paramelab} for a further discussion on parameter choices). The simulations are performed using the Piece-Wise Parabolic \citep[PPM;][]{1984JCoPh..54..174C} Lagrangian remap hydrodynamics code \texttt{VH-1}\footnote{\texttt{http://wonka.physics.ncsu.edu/pub/VH-1/}}.  

The key challenge in such instability simulations is a proper treatment of the radiation line-force in a time-dependent, highly structured stellar wind outflow. As the instability acts on small spatial length scales expensive non-local line-force integrations have to be carried out each hydrodynamical time-step. It is important that such non-local line-force computations should still approximate well steady-state Sobolev models in order to be consistent. To this end we rely on the Smooth Source Function (SSF) formulation \citep{1991ASIC..341..235O} that has been extensively described by OP96. 

All our model simulations start from a smooth, relaxed CAK wind model covering a non-uniform grid $(1-10R_\star)$ of 2000 cells. We apply a grid stretching factor s between subsequent radial cells $i$ and $i+1$ such that $s \equiv \Delta r_{i+1}/\Delta r_i = 1.002$. This allows the grid to start from a subsonic base that resolves the effective photospheric scale height. Inner boundary conditions are chosen such that the density $\rho_0$ is fixed and we set velocity by constant slope extrapolation. The lower boundary density $\rho_0$ is about 5-7 times the density at the sonic point \citep{2013MNRAS.428.1837S}. At the (supersonic) outer boundary all hydrodynamical variables are set by simple extrapolation assuming constant gradients.

\subsection{Radiation line-driving}\label{sec:winddriving}

For our wind modelling we assume driving by an ensemble of lines described by a power law line-distribution, as originally proposed by CAK, in line-strength $q$ according to the formalism of \citet{1995ApJ...454..410G}. Additionally, following OP96, we assume an exponential truncation of this line-distribution limited to a maximum strength $Q_{\mathrm{max}}$
\begin{equation}\label{eq:linedist}
\frac{dN}{dq} = \frac{1}{\Gamma(\alpha)\bar{Q}} \left( \frac{q}{\bar{Q}} \right)^{\alpha-2} e^{-q/Q_{\mathrm{max}}},
\end{equation}

\noindent with $\Gamma(\alpha)$ the Gamma function. We further describe the line-driven wind using $\alpha$, the CAK power law index denoting the relative contribution of the line-force from optically thick lines to the total line-force, and $\bar{Q}$ as a line-strength normalization factor describing the ratio of the total line-force to the electron scattering force if all lines were to be optically thin \citep{1995ApJ...454..410G}\footnote{Here we deviate from the OP96 ($\kappa$, $\kappa_0$, $\kappa_{\mathrm{max}}$) parametrization of the line-distribution, i.e.~our equation \eqref{eq:linedist}. Gayley's ($q$, $\bar{Q}$, $Q_{\mathrm{max}}$) has the advantage of being a dimensionless measure of line strength that is independent of a fiducial thermal speed dependence. The relation between our parametrization and OP96 follows from $\bar{Q} = \kappa_0 \varv_{\mathrm{th}} \Gamma(\alpha)^{1/(1-\alpha)}/(c\kappa_e)$.}. Regarding continuum opacities, we take, as is customary, an electron scattering contribution described by an opacity $\kappa_e =0.34$ cm$^2$ g$^{-1}$. Electron scattering effectively lowers gravity and is commonly captured by Eddington's Gamma $\Gamma_e \equiv \kappa_e L_\star/(4\pi GM_\star c)$. The effective escape speed at the stellar surface becomes then $\varv_{\mathrm{esc,eff}} = \sqrt{2GM_\star(1-\Gamma_e)/R_\star}$. This setup allows then for a computation of the average mass-loss rate $\langle \dot{M} \rangle$ and average terminal wind speed $\langle \varv_\infty \rangle$ in our simulations. 

We include a photospheric linear Eddington limb darkening description as motivated both by perturbation analysis and numerical simulations \citep{1985ApJ...299..265O,2013MNRAS.428.1837S}. As studied extensively in the latter paper, this leads to a net instability also at the stellar surface and so to structure formation somewhat closer to the photosphere than in simulations assuming a uniformly bright stellar disk.

We refer the reader to Appendix \ref{sec:appendix} for a full description on how the perturbed line-force is numerically computed.

%--------------------------------------------------------------------
\section{Numerical results}

\subsection{On the theoretical mass-loss rate predictions}  \label{sec:paramelab}
   
It is important to point out that the bi-stability jump as predicted from line-driven wind models \citep{1999A&A...350..181V} implies an increase in mass-loss rate most directly when considering models with fixed mass and luminosity for which effective temperature is varied. Given that the internal parameter space of B supergiants is very scattered in fundamental stellar parameters like $L_\star$, $M_\star$, and $R_\star$ \citep{2007A&A...467.1265B,2008A&A...478..823M,2018A&A...614A..91H} we do not consider such fixed luminosity-mass models. Instead, we consider models where we select an effective temperature and choose a stellar radius that is typical in the observed range of OB supergiants. From this radius and effective temperature a stellar luminosity is determined. Similarly we pick a reasonable stellar mass for each model based on observations of OB supergiants. 

We should also point out that our non-linear simulations do not fundamentally predict global mass-loss rates. We choose our model wind line-force parameters \citep[e.g.][]{2000A&AS..141...23P} such that our relaxed CAK simulation mass-loss rates are approximately the same as those predicted from the \citet{2001A&A...369..574V} recipe. Evolving such relaxed CAK models including the non-linear instability then provides average mass-loss rates $\langle \dot{M} \rangle$ quite comparable to those of \citet{2001A&A...369..574V}. Similarly, we extract average terminal wind velocities $\langle \varv_\infty \rangle$ from our simulation and compute the ratio to the effective escape speed. The retrieved simulation quantities are shown in Table \ref{table:simparams} for comparison. 

Inspection of Table \ref{table:params} shows that we have not computed any models that are part of the predicted bi-stability jump region of \citet{1999A&A...350..181V}. The reason for this is as outlined above, i.e.~our steady-state CAK models are required to converge to mass-loss rates close to those predicted by \citet{2001A&A...369..574V} using a realistic set of wind parameters. However, the \citet{2001A&A...369..574V} mass-loss rate recipe breaks down in the bi-stability region such that there is no clear prediction for what should actually happen here. Therefore, it is hard to assess and justify the correctness of any initial, steady-state Sobolev model and as such any line-driven instability simulation and wind clumping prediction within the bi-stability.

To that end our non-linear simulations do not provide a direct test of the predicted increase in mass-loss rates across the bi-stability jump as found by \citet{1999A&A...350..181V}. However, our simulations can provide evidence of whether wind clumping properties change between O and B supergiants, which affects mass-loss rates derived from observational diagnostics. The observational lack of evidence for increased mass-loss rates might then be explained in terms of applying incorrect clumping factors for OB supergiants across the bi-stability jump.

\begin{table}
\caption{Wind parameters obtained from instability simulations using the models of Table \ref{table:params}.}             % title of Table
\label{table:simparams}      % is used to refer this table in the text
\centering                          % used for centering table
\begin{tabular}{c c c c}        % centered columns (4 columns)
\hline\hline                 % inserts double horizontal lines
Model & $\langle \dot{M} \rangle$ & $\langle \varv_{\infty} \rangle$ &   $\langle \varv_{\infty} \rangle / \varv_{\mathrm{esc},\mathrm{eff}}$ \\    % table heading 
            & [$M_\odot$ yr$^{-1}$] &  [km s$^{-1}$] & \\
\hline                        
   OSG1 & $3.6\times 10^{-6}$ & 2297   & 3.4 \\    
   OSG2 & $2.0\times 10^{-6}$ & 1859   & 3.1 \\
   BSG1 & $9.3\times 10^{-7}$  & 769     & 1.9 \\
   BSG2 & $2.5\times 10^{-7}$ & 722      & 1.7 \\
\hline                                   
\end{tabular}
\end{table}
   
\subsection{Structure formation in O and B supergiant winds}
   
We illustrate first the basic, overall dynamics of an unstable 1D line-driven wind. Figure \ref{FIGstructureformation} shows the basic structure resulting from a non-linear LDI simulation, taken after a long enough simulation time which we set here by evolving over sufficient dynamical timescales $\tau_{\mathrm{dyn}} = R_\star /\varv_\infty $. This 1D snapshot illustrates the typical formation of (strong) shocks that compress matter into dense, spatially narrow clumps (really shells in 1D) separated by a high-speed rarefied medium. The models shown start from a relaxed CAK model and then evolve up to a point where the wind reaches a statistically steady flow. 
     
        \begin{figure}[!h]
   \centering
   \includegraphics[width=\hsize]{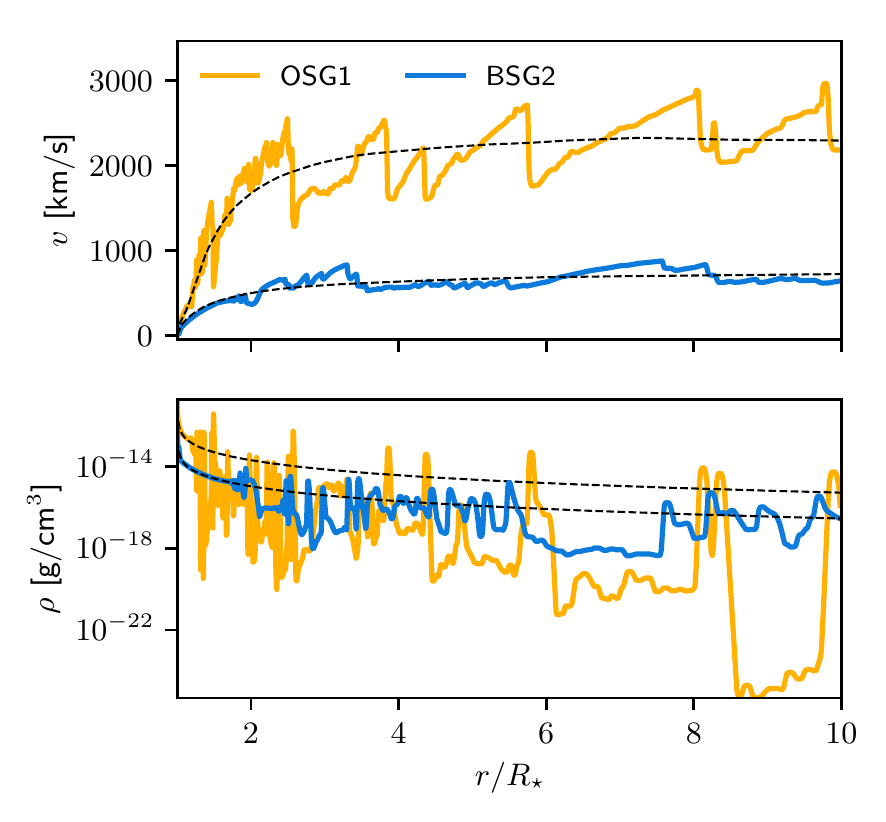}
      \caption{Velocity and density evolution of LDI in the wind of OB supergiants. Comparison is between our hottest and coolest model. The dashed line represents the average wind velocity/density profile.}
         \label{FIGstructureformation}
   \end{figure}
   
In Figure \ref{FIGstructureformation} we compare our hottest O supergiant model (OSG1) with that of the coolest B supergiant (BSG2). The figure directly shows the structural changes that occur in shock formation and resulting clump formation, as predicted by linear perturbation analysis in the previous section (see also Figure \ref{FIGgrowthrates}). Not only the strength of structure formation is different, but also its onset, i.e.~the dynamical timescale between O and B supergiants are quite different. A direct computation shows that for the O supergiant $\tau_{\mathrm{dyn}}\approx 14$ ks and for the B supergiant $\tau_{\mathrm{dyn}}\approx 65$ ks, meaning that B supergiants take longer to relax into a statistically steady-state flow. 
   
Figure \ref{FIGonset} shows a 2D contour of wind density $\rho$ and wind velocity $\varv$ over time of all our models. As motivated in the next section we concentrate here on the inner wind region where the H$\alpha$ line is formed ($r\leq 2R_\star$). All simulations extend to 300 ks for a direct comparison, but this only captures the start of structure formation in the B supergiant wind models due to the longer dynamical timescale. In density space (top panels) there is a significant contrast between the O and B supergiant winds. Specifically, the B supergiant winds display much lower density and velocity contrasts between high-density clumps (shells) and the rarefied regions in between them, as can be directly seen in the lower panel of Figure \ref{FIGstructureformation}. A O supergiant has regions of very rarefied gas compared to a B supergiant. Indeed, gas material is much more accelerated in a O supergiant wind than in a B supergiant wind (bottom panels). 

In accordance with our derived scaling relation above, the instability sets in much faster for the O supergiants than for the B supergiants. The O supergiants starts to form small-scale structures around 10 ks while the B supergiants only exhibits unstable behaviour around 50 ks. This result comes quite close to prediction made from equation \eqref{eq:scalingrel}, i.e.~computing $\Omega^{-1}$ shows that the actual development of non-linear structure starts on the order of several ks to several 10 ks for O and B supergiants, respectively. If we let these simulations run for sufficient dynamical timescales, re-occurring periodic structures appear in density and velocity space. This is already clearly visible in the O supergiant winds while periodic structures still have to develop in the B supergiant winds. Essentially, limb-darkening lowers scattering contributions in the transonic wind region such that the wind solution becomes of unstable nodal type instead of stable X-type. The nodal solution topology branch has a degenerate set of shallow-slope solutions which are reached in semi-regular intervals of the simulation \citep[see][for more quantitative details]{2015MNRAS.453.3428S}.
   
    \begin{figure*}[!h]
   \centering
   \resizebox{.95\hsize}{!}
   	 {\includegraphics{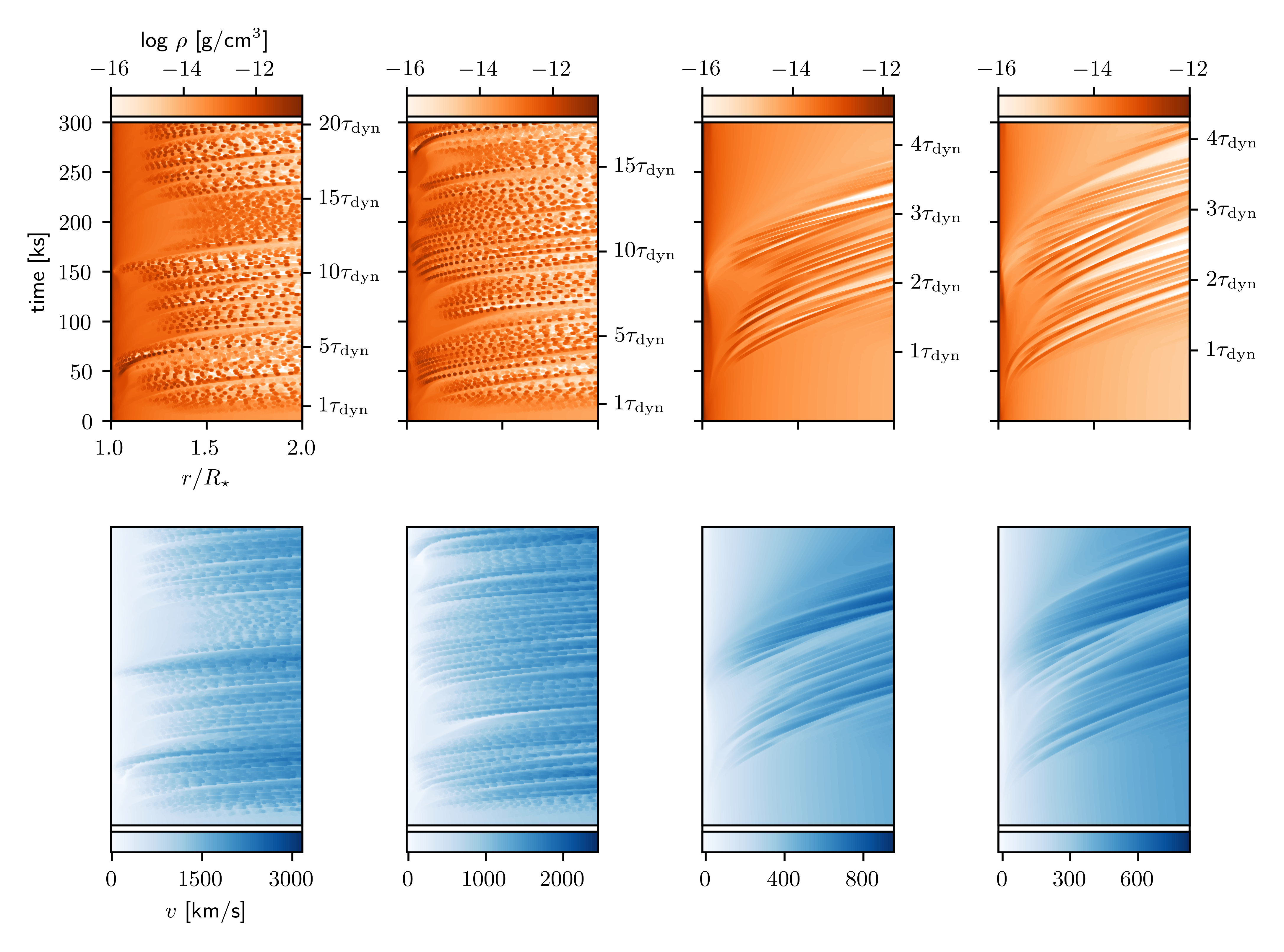}}
      \caption{(Top row) Logarithmic wind density structure From left to right: OSG1, OSG2, BSG1, BSG2. (Bottom row) Wind velocity structure of the corresponding models. The rarefied medium in the density plot is characterized by high velocities while the dense, clumps move at much lower velocities. All models extent to 300 ks but the B supergiant models actually run for a much longer time (see text). Also illustrated is the difference in dynamical timescales that governs the onset of structure formation.}
         \label{FIGonset}
   \end{figure*}      
      
\subsection{Wind clumping across the bi-stability jump}
   
Having discussed the basic structures of the non-linear evolution of the LDI we here present statistically computed properties for our OB supergiant models. The different dynamical properties between O and B supergiants also alter the wind clumping properties of the wind. To describe the clumping properties we use a clumping factor
\begin{equation}
f_{\mathrm{cl}} = \frac{\langle \rho^2 \rangle}{\langle \rho \rangle^2},
\end{equation}

\noindent with brackets denoting time-averages after a statistically steady flow has developed. Essentially, the O and B supergiant models are all averaged between 10 - 25$\tau_{\mathrm{dyn}}$ which translates to about 100 - 300 ks for the O supergiant models and about 700 ks - 2 Ms for the B supergiant models. Following this approach we take into account the difference in dynamical timescale and make all the simulations insensitive to initial conditions.   
   
The left panel of Figure \ref{FIGfcl} shows the relative change of the radially varying wind clumping factor $f_{\mathrm{cl}}$ between our models. The middle and right panels show average wind clumping factors over the full wind and inner wind $(1-2R_\star)$, respectively. We compute these averages for each of our models and show them as a function of effective temperature to illustrate the behaviour across the bi-stability jump. Already the average over the full wind shows a significant difference between wind clumping properties on the hot and cool side of the bi-stability jump. As the inner wind is the region where the H$\alpha$ line is being formed, it is important to consider a wind clumping average here. The H$\alpha$ line is a prime spectral diagnostic used to derive empirical mass-loss rates across the bi-stability jump \citep[e.g.][]{2008A&A...478..823M}. However, such H$\alpha$ fitting also depends sensitively on the wind clumping properties, due to the $\rho^2$-dependence (see Introduction). Effectively, when deriving $\dot{M}$ from H$\alpha$ fitting, one is really deriving the product $\dot{M} \sqrt{f_{\mathrm{cl}}}$. This immediately shows a degeneracy occurs when fitting line-profiles because different combinations of $\dot{M}$ and $f_{\mathrm{cl}}$ can provide the same invariant $\dot{M} \sqrt{f_{\mathrm{cl}}}$.

   \begin{figure*}[!h]
   \resizebox{.95\hsize}{!}
            {\includegraphics{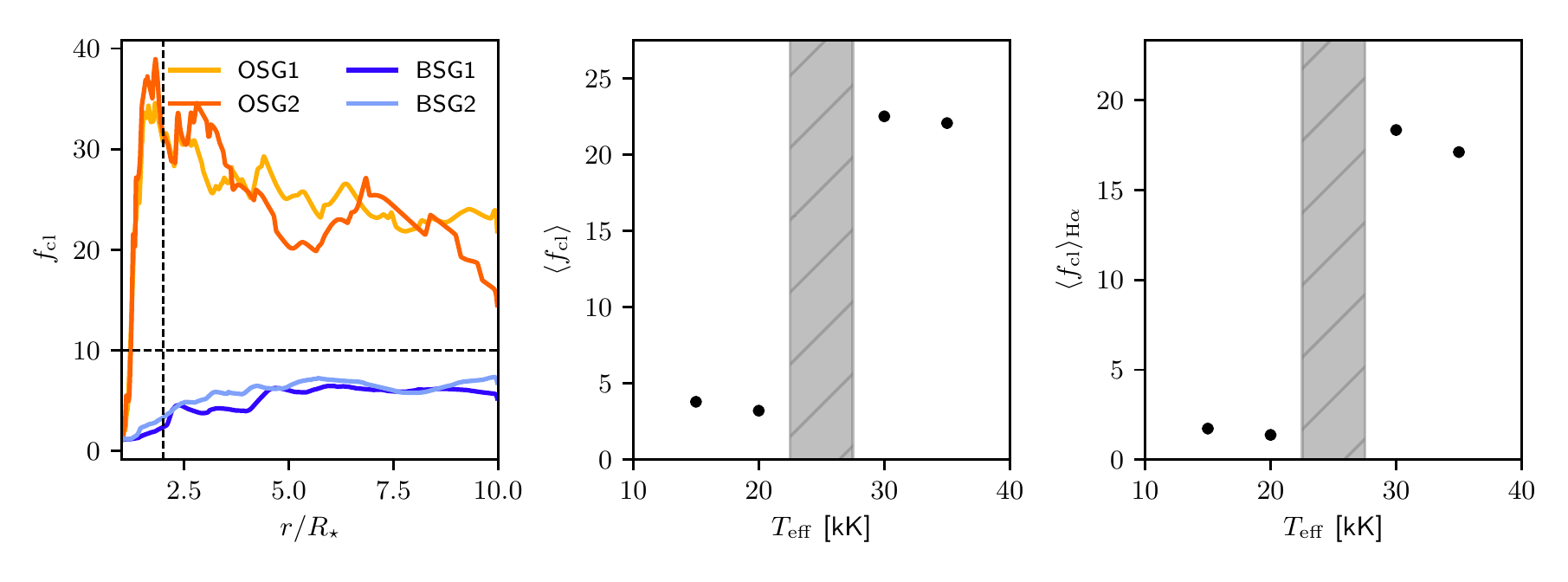}}
           \caption{(Left panel) Relative change in wind clumping between our models. The dashed lines mark the inner wind where H$\alpha$ line-formation occurs and the typical value  $f_{\mathrm{cl}} = 10$ often used in diagnostic modelling. (Middle panel) Averaged clumping factor over the whole wind. (Right panel) Averaged clumping factor in the H$\alpha$ region. The grey area represent the bi-stability region as predicted by \citet{1999A&A...350..181V}.}
         \label{FIGfcl}
   \end{figure*} 
   
On the hot side of the bi-stability jump, our simulations show $f_{\mathrm{cl}}\sim 17$ in the H$\alpha$ forming region. On the other hand, the cool side of the bi-stability jump shows much lower factors $f_{\mathrm{cl}}\sim 2$.  Note that our predicted wind clumping factors at the hot side of the bi-stability jump agree reasonably well with typical values inferred from spectral diagnostical studies \citep{2011A&A...535A..32N,2013MNRAS.428.1837S}. This makes our study quite robust in a relative sense, even though there are still many uncertainties related to a quantitative treatment of the LDI \citep[e.g.][]{2018A&A...611A..17S}. At the cool side we find very low $f_{\mathrm{cl}}\approx 2$, suggesting that it is important to account correctly for the different wind clumping properties across the bi-stability jump. From a differential viewpoint, a key result here is thus that our models suggest that the winds of O and B supergiants are differently structured in terms of clumping.

The lack of structure formation and corresponding shocks is also important for intrinsic X-ray emission of B supergiant winds. A key observational result is that the X-ray flux drops in B star winds as compared to their O star counterparts \citep{1994ApJ...421..705C,1997A&A...322..167B,2009A&ARv..17..309G}. This is in agreement with our simulations, although we are here not computing the shock-heated regions directly. To facilitate a discussion around shocks, we define a velocity dispersion
\begin{equation}
\varv_{\mathrm{disp}} = \sqrt{\langle \varv^2 \rangle - \langle \varv \rangle^2},
\end{equation}

\noindent which is a measure of the standard deviation of wind velocity and describes the typical velocity around the mean wind flow speed (brackets denote again a time-average). In the absence of direct X-ray modelling the velocity dispersion is a reasonable proxy for the overall intrinsic X-ray emission expected from LDI shocks.

Figure \ref{FIGdispersion} shows the velocity dispersion computed for our OB supergiant models. Clearly there is a large difference between the velocity dispersion in the models. For the O supergiant wind a steep increase in velocity dispersion results from the initial strong amplification of small-scale velocity structures near the stellar base by the LDI. Eventually these small-scale velocity structures situate themselves in the outer wind where almost no further amplification occurs. It is, therefore, not surprising to see that the B supergiant wind has neither a steep initial increase nor achieves large values in velocity dispersion.

   \begin{figure}
   \centering
   \includegraphics[width=\hsize]{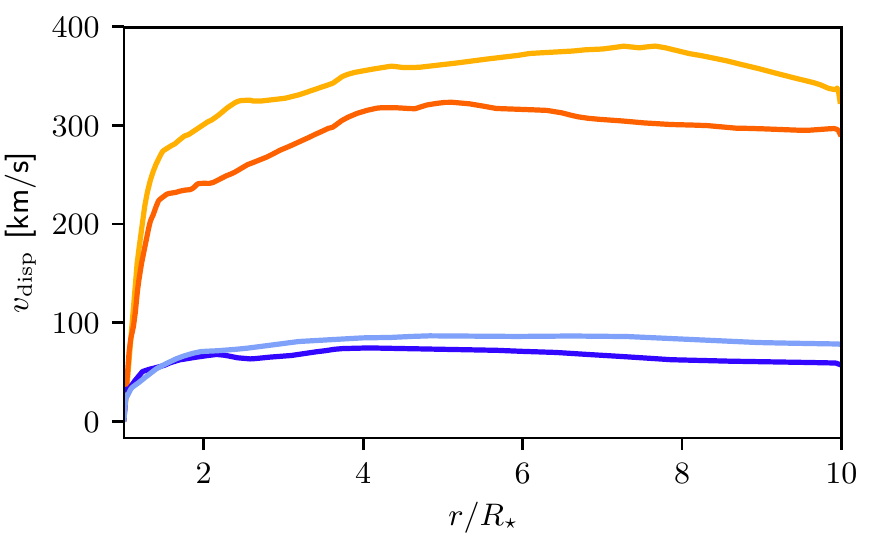}
      \caption{Velocity dispersion inside the wind computed from our simulations of OB supergiant models.}
         \label{FIGdispersion}
   \end{figure}  
   
Typical shock jump velocities in O supergiant winds are on the order of $\varv \sim 300$ km s$^{-1}$ giving rise to shock temperatures on the order of $T \sim 1$ MK. Simple visual inspection of Figure \ref{FIGdispersion} shows our O supergiant models have $\varv_{\mathrm{disp}} \sim 300$ km s$^{-1}$ which is enough to produce intrinsic X-ray emission, whereas the velocity dispersion in a B supergiant model is several factors lower ($\varv_{\mathrm{disp}} \sim 70$ km s$^{-1}$), and so will not result in much (or any) shock-heated gas hot enough to emit in the X-ray band. Physically this is a combined consequence of the weaker clumping properties and lower wind speeds of B supergiant line-driven winds (see also previous figures). 

Finally, we also note that lower levels of velocity dispersion in B supergiant winds than in O supergiant winds are generally consistent with quantitative spectroscopic modelling of saturated UV P-Cygni lines. Namely, in such diagnostic modelling the velocity dispersion is conventionally modelled using a so-called `wind microturbulence', which in order to match observations typically is assumed to scale with the local wind speed, or reaching maxima on the order of $0.1 \varv_\infty$ \citep[e.g.][]{1993A&A...279..457P}. Since B supergiant winds have much lower $\varv_\infty$ than O supergiant winds, this agrees well with our findings here.

%--------------------------------------------------------------------
\section{Conclusions and Future Work}

Central in this work is that theoretical predictions of clumping in O and B supergiant winds differ significantly from each other. With the help of analytical perturbation analysis we computed a scaling relation for the linear growth rate of the LDI for a single optically thick line. This calculation shows that there is a significant difference in instability growth rates between O and B supergiants. Likewise we find that the onset of structure formation occurs at different times. We confirm these analytical findings using time-dependent line-driven instability simulations that describe the non-linear evolution of the LDI.

Extending these line-driven instability simulations to B supergiants has allowed us to study the behaviour of wind clumping across the predicted bi-stability jump in massive star winds \citep{1990A&A...237..409P,1995ApJ...455..269L,1999A&A...350..181V}. A key result of our study is that simulations show a significant difference in wind clumping properties between the hot and cool side of the bi-stability jump. Therefore, this suggests that accounting correctly for wind clumping is important in deriving empirical mass-loss rates across the bi-stability jump, \emph{if} they are derived by means of the $\rho^2$-sensitive H$\alpha$ line. Namely, in these diagnostic H$\alpha$ studies of mass loss across the bi-stability jump, clumping effects are typically neglected entirely or treated as a constant across the bi-stability jump \citep[e.g.][]{2008A&A...478..823M}. In turn, this may then also help alleviate current problems with massive star evolution models that do not assume a mass loss increase over the bi-stability jump \citep{2010A&A...512L...7V,2017A&A...598A...4K}. 

Building on the results of this paper, it would next be interesting to study in more detail line-driven instabilities across the \emph{internal} parameter space of B supergiants. The stellar parameter space of B supergiants is very scattered \citep{2007A&A...467.1265B,2008A&A...478..823M,2018A&A...614A..91H} making it a natural question to what extent such different sets of stellar parameters also affect clumping properties in the wind. Along these lines it would then also be interesting to consider models at fixed luminosity and mass like those of \citet{1999A&A...350..181V}. As directly evident from the linear scaling relation, equation \eqref{eq:scalingrel}, we indeed expect that also such fixed luminosity-mass models would exhibit significantly weaker wind clumping across the bi-stability jump.
  
Since the modelled velocity dispersion is much lower in BSG than OSG winds, our simulations provide a theoretical rationale to observations showing that X-ray fluxes drop significantly in single B-star winds as compared to their O-star counterparts \citep{1994ApJ...421..705C,1997A&A...322..167B,2009A&ARv..17..309G}. This then also naturally explains the diagnostic finding that the wind microturbulence needed to fit observations of strong UV wind lines typically scales with the terminal speed. Moreover, the lack of strong shocks and structure formation might also affect wind accretion in High-Mass X-ray Binaries (HMXBs). These systems consist often of a B supergiant and an orbiting object that accretes material, e.g.~the prototype system Vela X-1 where the donor star is a early B supergiant with a slow and dense wind characteristic of a simulation below the bi-stability jump \citep{2018A&A...610A..60S}. However, in the accretion models by \citet{2018MNRAS.475.3240E} targeting Vela X-1, a multi-D LDI simulation of an O supergiant was used to simulate the clumpy wind accretion and the corresponding time-variation in X-ray luminosity. Preliminary calculations made by us suggest that also in such multi-D models the structure formation in B supergiants is still much weaker than in O supergiants. Consequently, the amount of LDI generated accretion rate variations may be somewhat overpredicted in the models of \citet{2018MNRAS.475.3240E}. In a follow-up work, we plan to study in detail the nature of multi-D LDI models of B supergiants on the accretion rate and corresponding variations of X-ray luminosities in such HMXBs. 

Finally, we point out that this work is a first theoretical study of the behaviour of wind clumping on both sides of the bi-stability jump. Our primary aim is to point out the difference in wind clumping properties in the O and B supergiant regimes. We show how the wind clumping properties change across the bi-stability jump and discuss whether this may provide an explanation to the lack of observational evidence for increasing mass-loss rates across the bi-stability jump. A quantitative wind clumping description depends on the treatment of limb-darkening, photospheric base perturbations, the source function, and the amount of lateral fragmentation occurring in (much more complex) multi-dimensional wind clumping models \citep{2003A&A...406L...1D,2018A&A...611A..17S}.

Moreover, our simulations make only predictions of clumping factors, but not the overall stellar mass-loss rate. Extending this study to also treat stars within the bi-stability jump region must be left to future work once more reliable predictions for the smooth, steady-state winds in this domain are in place. However, we have calibrated our 1D models so that they approximately agree with observed clumping factors on the hot side of the bi-stability jump. This means that, even though there are still some significant uncertainties regarding the input physics used to model the LDI, the simulations presented here should be quite robust in a relative sense. As such, this paper provides a good first study of the differential effect of theoretical predictions of wind clumping of O and B supergiants across the bi-stability jump.

\begin{acknowledgements}
Many thanks towards the anonymous referee for raising helpful comments that improved the manuscript. The authors wish to thank Stan Owocki for insightful discussions. We also thank the KU Leuven {\sc equation}-group for their valuable input as well as weekly supply of cake. F.A.D. and J.O.S. acknowledge support by the Belgian Research Foundation Flanders (FWO) Odysseus program under grant number G0H9218N. N.D.K. acknowledges support from the KU Leuven C1 grant MAESTRO C16/17/007. 
\end{acknowledgements}


\bibliography{refs} % your references Yourfile.bib
\bibliographystyle{aa} % style aa.bst

\begin{appendix} %First appendix

\section{Computing the perturbed line-force}\label{sec:appendix}

As discussed in the Introduction, the Sobolev method is not suitable for computing the small-scale structures arising from the line-deshadowing instability as the instability acts on length scales smaller than the Sobolev length. This means that the line-force does not depend any more only on local flow quantities, but instead there is a non-local coupling between the radiation field and outflowing gas. Such full non-local calculations are computationally daunting and approximate expressions for the line-force have to be found. One general approximation is based on the escape integral probability method (OP96).

We define the escape probability for a single line, following the \citet{1995ApJ...454..410G} parametrization, as
\begin{equation}
b_q(\mu,r) \equiv \int_{-\infty}^{+\infty} dx \phi \left( x-\mu \varv(r)/\varv_{\mathrm{th}} \right) e^{-qt(x,p,z,z_b)},
\end{equation}

\noindent where $\phi$ is a profile function (taken to be Gaussian) at observer's frame frequency $x \equiv (\nu/\nu_0 - 1)c/\varv_{\mathrm{th}}$ defined from line-centre $\nu_0$ in terms of the frequency broadening from ion thermal motions, $\mu$ is a local direction cosine at radius $r$, and $qt$ is a frequency dependent optical depth for a line at strength $q$ 
\begin{equation}
qt(x,p,z_1,z_2) \equiv \int_{z_1}^{z_2} \kappa \rho(r') \phi \left( x-\mu \varv(r')/\varv_{\mathrm{th}} \right) dz'.
\end{equation}

We do here the integrations between two positions $z_1 < z_2$ along a certain ray $p$ (impact parameter) from the origin where $z\equiv \mu/r$ and $r^2 \equiv p^2 + z^2$. Notice that $\kappa$ in the above equation is not the usual opacity (expressed in units of [cm$^2$ g$^{-1}$]), but rather is a frequency integrated line-opacity (units [cm$^2$ g$^{-1}$ Hz$^{-1}$]). Specifically, $\kappa \equiv \chi / \rho$ where $\chi$ is a frequency integrated line-strength \citep[e.g. as defined in][see their equation 2]{1993A&A...279..457P}.

In the Sobolev approximation the velocity variation dominates the integral $(\varv \gg \varv_{\mathrm{th}})$ such that $\kappa \rho$ becomes approximately constant over a Sobolev length $L\equiv \varv_{\mathrm{th}}/(d\varv /dr)$, i.e.~the integral is analytic. For physics occuring on spatial scales smaller than $L$ (line-driven instabilities), this integral is inherently \emph{non-local} and requires numerical evaluation.

The stellar wind off a hot, luminous star is not driven by a single line but a large collection of lines. For an ensemble of lines modelled by a line-distribution (see Section \ref{sec:winddriving}, also definition in OP96) we define an escape probability
\begin{equation}\label{eq:ei}
\begin{split}
b(\mu,r) &\equiv \Gamma(\alpha)^{1/(1-\alpha)} \int_0^{+\infty} dq \frac{dN}{dq} \frac{q}{\bar{Q}} b_q(\mu,r), \\
              &= \Gamma(\alpha)^{(1+\alpha)/(1-\alpha)} \int_0^{+\infty} dq \frac{\phi \left( x-\mu \varv(r)/\varv_{\mathrm{th}} \right)}{\left( \bar{Q}t(x,p,z,z_b) + \bar{Q}/Q_{\mathrm{max}} \right)^\alpha}.
\end{split}
\end{equation}

Starting from the definition of the radiative line-acceleration it is possible to recast it in terms of the above ensemble escape probability (detailed derivation in OP96). The line-force expressions for the direct and diffuse radiation become
\begin{equation}\label{eq:gdir}
g_{\mathrm{dir}} = g_{\mathrm{thin}}  \frac{\langle \mu D(\mu, r) b(\mu, r) \rangle}{\langle \mu D(\mu, r) \rangle},
\end{equation}

\noindent and 

\begin{equation}\label{eq:gdiff}
g_{\mathrm{diff}} = -2g_{\mathrm{thin}} \frac{S(r)}{I_\star}\langle \mu b(\mu,r) \rangle,
\end{equation}

\noindent with $g_{\mathrm{thin}} \equiv \bar{Q}g_e$ the line-acceleration in the optically thin limit and $g_e = \kappa_e L_\star/(4\pi r^2 c)$ is the acceleration due to continuum electron scattering. Angle brackets in the line-force expressions denote average angle integrations. The total force due to radiation $g_{
\mathrm{rad}}$ is the sum of the above continuum force and line-forces, i.e. $g_{\mathrm{rad}} \equiv g_e + g_{\mathrm{dir}} + g_{\mathrm{diff}}$.

The function $D(\mu,r)$ contains any reductions of stellar intensity in case the star is not assumed to be a uniformly bright disk: $D(\mu,r) = 1$ $(\mu_\star < \mu < 1)$ and $D(\mu,r)=0$ ($\mu < \mu_\star$), where $\mu_\star  = \sqrt{1-(R_\star/r)^2}$ gives the angular size of the star. To account for the fact that the star is not a uniformly bright disk, we include such a $D(\mu,r)$ that describes a photospheric linear Eddington limb-darkening law \citep[][equation 17.17]{2014HubenyMihalas}
\begin{equation}\label{eq:eddld}
D(\mu, r) = \frac{1}{2} + \frac{3}{4}\sqrt{\frac{\mu^2-\mu_\star^2}{1-\mu_\star^2}},
\end{equation}

\noindent which has the additional effect of reducing the stabilizing effect of diffuse radiation near the stellar base.

Additionally, in order to compute the diffuse line-force in equation \eqref{eq:gdiff} we require an expression for the source function $S(r)$. The optically thin limit turns out to be a good approximation and the source function can be well approximated (OP96) by
\begin{equation}\label{eq:sf}
S(r) = \langle I_\star D(\mu, r) \rangle.
\end{equation}

Plugging equation \eqref{eq:eddld} into the above gives then \citep{2013MNRAS.428.1837S}
\begin{equation}
S^{\mathrm{LD}}(r) = \frac{I_\star}{16} \left( 7 - 4\mu_\star + 3\mu_\star^2 \frac{\ln\left[ \mu_\star/(1+\sqrt{1-\mu_\star^2} ) \right]}{\sqrt{1-\mu_\star^2}} \right)
\end{equation}

For convenience it is useful to cast the angle integration of the line-forces in terms of a ray parameter $y \equiv (p/R_\star)^2$ with direction cosine $\mu_y = \sqrt{1-y(R_\star/r)^2}$. Applying this ray formulation and the Eddington limb-darkened source function into the expressions for the line-forces, i.e.~equation \eqref{eq:gdir} and \eqref{eq:gdiff} give
\begin{equation}
g_{\mathrm{dir}}(r) = g_{\mathrm{thin}} \left( \frac{1}{2} \int_0^1 b(\mu_y,r)dy + \frac{3}{4} \int_0^1 \sqrt{1-y}b(\mu_y,r)dy  \right),
\end{equation}

\noindent and
\begin{equation}
g_{\mathrm{diff}}(r) = \frac{S(r)}{S^{\mathrm{LD}}(r)} \frac{g_{\mathrm{thin}}}{2(1+\mu_\star)} \int_0^1 \left( b_-(\mu_y,r) - b_+(\mu_y,r) \right) dy,
\end{equation}

\noindent in which the diffuse line-force follows a `two-stream' approximation of inward (`$-$', $\mu<0$) and outward (`$+$', $\mu>0)$ streaming photons. Additionally we correct with a factor $(r/R_\star)^2$ as the ray integration should formally extend to $y=(r/R_\star)^2$ \citep{1991ASIC..341..235O}. It has been shown that $y=0.5$ is a good ray parameter choice and approximates the full stellar disk integration within a few percent for most of the wind \citep{2013MNRAS.428.1837S}.

The outward escape probability $b_+(\mu_y,r)$ is computed according to equation \eqref{eq:ei} using an outward frequency dependent optical depth $t_+ \equiv qt(+x,y,r)$. The inward escape probability $b_-(\mu_y,r)$ requires a corresponding inward frequency dependent optical depth $t_-$ which can be computed from $t_+$ by using (OP96)
\begin{equation}
t_-(-x,y,R_{\mathrm{max}}) = t_+(2\mu_y \varv(R_\mathrm{max})/\varv_{\mathrm{th}} - x, y,R_{\mathrm{max}}).
\end{equation}

With the above equations, the line-forces are computed at each hydrodynamical time-step and added together to get the total non-local line-force acting on the outflowing stellar wind.

\end{appendix}

\end{document}